\documentclass{ws-procs975x65}

\begin{document}

\title{Monte Carlo Simulations of Opinion Dynamics}

\author{S. Fortunato}

\address{Fakult\"at f\"ur Physik, Universit\"at Bielefeld,\\
D-33501, Bielefeld, Germany \\
and \\
Dipartimento di Fisica e Astronomia and INFN sezione di Catania,\\
Universit\'a di Catania,\\
Catania I-95123, Italy\\
E-mail: fortunat@physik.uni-bielefeld.de}

\maketitle

\abstracts{We briefly introduce a new promising field of applications 
of statistical physics, opinion dynamics, where the systems at study 
are social groups or communities and 
the atoms/spins are the individuals (or agents) belonging to such groups. The opinion
of each agent is modeled by a number, integer or real, and 
simple rules determine how the opinions vary as a consequence of discussions 
between people. Monte Carlo simulations of consensus models lead to patterns of 
self-organization among the agents which fairly well reproduce the trends observed
in real social systems.}

\section{Introduction}

Statistical physics teaches us that, even when it is impossible to foresee 
what a single particle will do, one can often 
predict how a sufficiently large number of particles will behave, in spite of the 
eventually large differences between the variables describing the state of the individual particles.

This principle holds, to some extent, for human societies too.
It is nearly impossible to predict when one person will die, as the death depends on
many factors, most of which are hard to control: nevertheless 
statistics of the mortality rates of large populations 
are stable for long times and have been studied for over three centuries.
We then come to the crucial question: 

\vskip0.4cm

\centerline{\it Can one describe social behaviour through statistical physics?}
 
\vskip0.4cm

The question is tricky, and bound to trigger hot debates within the physics community.
On the one hand, society is made of many individuals which interact mostly locally 
with each other, like in classical statistical mechanical systems. On the other hand,  
social interactions are not mechanical and are hardly reproducible. However we expect that
the aspects of collective behaviour and self-organization in a society may be reasonably
well described by means of simple statistical mechanical models and by now 
several such models have been introduced and analyzed, giving rise to the new field 
of sociophysics\cite{callen,galam,weidlich}. 

In this contribution we shall concentrate on opinion dynamics.
The spread and evolution of opinions in a society has always been a central 
topic in sociology, politics and economics. One is especially interested in understanding
the mechanisms which favour (or hinder) the agreement among people of different 
opinions and/or the diffusion 
of new ideas. 

Early mathematical models
of opinion dynamics date back to the $50$'s, but the starting point for
quantitative investigations in this direction is marked by the theory of social impact 
proposed by Bibb Latan{\'e}\cite{lata}. The impact is a measure of the 
influence exerted on a single individual by those agents which interact with him/her (social neighbours).
Models
based on social impact\cite{schw} were among the 
first microscopic models of opinion dynamics. They are basically
cellular automata, where one starts by assigning, usually at random,
a set of numbers to any of the $N$ agents of a community.   
One of these numbers is the opinion, the others describe specific features of the agents, like 
persuasiveness, supportiveness, tolerance, etc. Society is modeled as a graph, and 
each agent interacts with its geometric neighbours, which represent friends or close relatives.
The procedure is iterative: at each iteration 
one takes a set of interacting agents and updates their opinions (or just the opinion
of a single agent), according to a simple dynamical rule. After many iterations, the system usually 
reaches a state of static or dynamic equilibrium, where the distribution of the 
opinions among the agents does not change shape, even if the agents themselves still change their mind.
The dynamics usually favours the agreement of groups of agents about the same opinion, so that 
one ends up with just a few opinions in the final state. In particular it is 
possible that all agents share the same opinion (consensus), 
or that they split in two or more factions. 

Most results on opinion dynamics derive from Monte Carlo simulations of the 
corresponding cellular automata.
We shall here shortly present two 
basic consensus models: the Bounded Confidence Model (BCM)\cite{HK,D}
and the Sznajd Model\cite{Sznajd} (SM). 
For a complete exposition of the recent results on these models we refer to\cite{stauf0,stauf3}.  
Due to lack of space we 
are forced to omit the discussion of other important classes of opinion dynamics, like 
the voter models\cite{voter}, the majority rule models\cite{galam1} and the 
Axelrod model\cite{axel}. 

\section{The Bounded Confidence Model}

The BCM is based on the simple consideration that two persons usually discuss 
with each other about a topic only if their opinions on that topic are quite close to each 
other, otherwise they quarrel or avoid discussing. This can be easily modeled by introducing a parameter 
$\epsilon$, called confidence bound, and by checking whether the opinions $s_i$ and $s_j$
of two social neighbours $i$ and $j$ differ from each other by less than $\epsilon$. If this 
were the case, we say that the opinions of the two agents are compatible and 
they can start a conversation which may lead to variations of their 
opinions. Opinions can be integers or real numbers; the 
opinions are initially distributed at random among the agents.
The number of opinion clusters $n_c$ in the final configuration depends on the confidence bound:
if $\epsilon$ is small, $n_c$ is roughly $1/\epsilon$; above some threshold $\epsilon_c$
the system attains consensus.   
There are two main versions of the BCM, which are characterized by two different
dynamical rules of opinion updating: the consensus model of Deffuant et al.\cite{D} (D)
and that of Krause-Hegselmann\cite{HK} (KH). Here we shall discuss the latter.

\subsection{Krause-Hegselmann}

\begin{figure}[hp] 
\centerline{\epsfxsize=3.4in\epsfbox{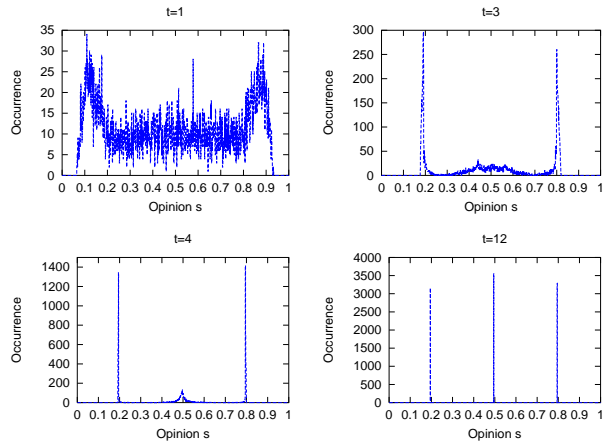}}   
\caption{Time evolution (in Monte Carlo steps per agent) of the opinion distribution of the KH model 
for a society where everybody talks to everybody else. The number of agents is 
$10000$,  $\epsilon=0.13$. 
The agents form three different factions in the final state.
\label{fig1}}
\end{figure}

The iteration of the KH model in the case of real-valued 
opinions consists of the following three steps:

\begin{enumerate}
\item{An agent $A$ is selected, sequentially or at random;}
\item{One checks which of the neighbours of A have opinions compatible with that of A.}
\item{The new opinion of A is the average of the opinions of its compatible neighbours.} 
\end{enumerate}

The dynamics of the model is not trivial because the opinion space is bounded
(typically [0, 1]): in fact, the inhomogeneities at the edges determine 
density variations in the opinion distribution, which propagate towards the center (Fig. \ref{fig1}).

For integer opinions, the update rule is even simpler\cite{santo}: agent $A$ takes the opinion
of one of its compatible neighbours, chosen at random.   
This rule recalls that of the voter\cite{voter} and Axelrod\cite{axel} models. In a society where
everybody talks to everybody else, if there are $Q$ 
possible choices for the agents and the condition of compatibility
for two opinions $S_i$ and $S_j$ is $|S_i-S_j|\leq 1$, the 
community always reaches consensus provided $Q\leq 7$.

\section{The Sznajd Model}

The SM is probably the most studied consensus model of the last years. The reasons of its 
success are the intuitive ``convincing rule'' and the deep relationship with 
spin models like Ising. One starts with a simple remark: an individual is more easily 
convinced to change its mind if more than just a 
single person try to persuade him/her. So, if two or more of our friends 
share the same view about some issue, it is likely that they will convince us to accept that 
view, sooner or later. 

\begin{figure}[hp] 
\centerline{\epsfxsize=3.2in\epsfbox{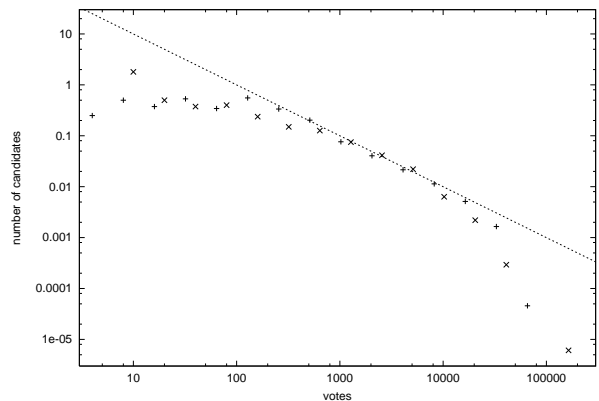}}   
\caption{Histogram of the fraction of candidates receiving a given number of votes 
for 1998 election in the state of Minas Gerais (Brazil). 
A simple election model based on Sznajd opinion 
dynamics reproduces well the central pattern of the data.
The data points are indicated by $\times$, the results of the election model by $+$ (from Ref. 15).
\label{fig2}}
\end{figure}

In the most common implementation of the model, a group of neighbouring agents which happen
to share the same opinion imposes this opinion to all their neighbours. The ``convincing'' pool 
of friends can be a pair of nearest-neighbours on a graph, or groups of three or more 
neighbours like triads on networks or plaquettes on a lattice. 
One usually starts from a random distribution of opinions among the agents, with a fraction $p$ 
of agents sharing the opinion $+1$ (the rest of the agents 
having opinion $-1$). In the absence of perturbing factors like noise, the state of the system
always converges towards consensus and a phase transition is observed as a function of the 
initial concentration $p$: for $p<1/2$ ($>1/2$) all agents end up with opinion $-1$ ($+1$). 

Since the original formulation of the model\cite{Sznajd}, for a one-dimensional chain of agents, countless
refinements have been made, which concern the type of graph, the updating rule, the introduction of 
external factors like a social temperature, advertising and ageing, etc. (for more details see\cite{stauf0,stauf3}). 

The Sznajd dynamics has been used to devise simple election models
which reproduce the bulk behaviour of votes distributions of real elections\cite{bern,gonz} (Fig. \ref{fig2}):  
this is at present the strongest validation of the SM.

\section{Conclusions}

Sociophysics and in particular opinion dynamics are moving their first steps, and there is still
a lot to do. Nevertheless the first results are encouraging and the hope to 
explain in this way the collective behaviour of social systems is strong. For the future 
it is necessary to gather more data from real systems and to open collaborations with sociologists.

\section*{Acknowledgments}

I thank D. Stauffer for letting me discover this fascinating field
and the Volkswagen Foundation for financial support.

\end{document}